\documentstyle[12pt]{article}
\newcommand{\sect}[1]{\setcounter{equation}{0}\section{#1}\indent}
\renewcommand{\theequation}{\thesection.\arabic{equation}}
\renewcommand{\thefootnote}{\fnsymbol{footnote}}
\newcommand{\EQ}{\begin{equation}}
\newcommand{\EN}{\end{equation}}
\newcommand{\bea}{\begin{eqnarray}}
\newcommand{\ena}{\end{eqnarray}}
\newcommand{\vs}[1]{\vspace{#1 mm}}

\newcommand{\pa}{\partial}

\newcommand{\uda}{\nearrow \kern-1em \searrow}

\newcommand{\la}{\lambda}
\newcommand{\La}{\Lambda}

\newcommand{\ti}{\tilde}
\newcommand{\th}{\theta}

\makeatletter
\def\eqnarray{%
 \stepcounter{equation}%
 \let\@currentlabel=\theequation
 \global\@eqnswtrue
 \global\@eqcnt\z@
 \tabskip\@centering
 \let\\=\@eqncr
 $$\halign to \displaywidth\bgroup\@eqnsel\hskip\@centering
 $\displaystyle\tabskip\z@{##}$&\global\@eqcnt\@ne
 \hfil$\displaystyle{{}##{}}$\hfil
 &\global\@eqcnt\tw@$\displaystyle\tabskip\z@{##}$\hfil
 \tabskip\@centering&\llap{##}\tabskip\z@\cr}
\makeatother

\begin{document}

\begin{titlepage}
\setcounter{page}{0}
\begin{flushright}
EPHOU 98-009\\
July 1998\\
hep-th/9807062\\
\end{flushright}

\vs{6}
\begin{center}
{\Large Prepotentials, Bi-linear Forms on Periods and 
Enhanced Gauge Symmetries in Type-II 
Strings
}

\vs{6}
{\large
Takahiro
Masuda
\footnote{e-mail address: masuda@particle.sci.hokudai.ac.jp}
\\ and \\
Hisao Suzuki\footnote{e-mail address: hsuzuki@particle.sci.hokudai.ac.jp}}\\
\vs{6}
{\em Department of Physics, \\
Hokkaido
University \\  Sapporo, Hokkaido 060 Japan} \\
\end{center}
\vs{6}

\centerline{{\bf{Abstract}}}
  We construct a bi-linear form on the periods of Calabi-Yau spaces. These are used to obtain the  prepotentials  around conifold singularities in type-II strings compactified on Calabi-Yau space. 
The explicit construction of the bi-linear forms is achieved  for the  one-moduli models as well as two moduli models with K3-fibrations where the enhanced gauge symmetry is known to be observed at conifold locus. We also 
show how these bi-linear forms are related with the existence 
of flat coordinates. 
 We list the resulting prepotentials in two moduli models around 
the conifold locus, which contains  $\alpha'$ corrections of 4-D 
N=2 SUSY $SU(2)$ Yang-Mills theory as the stringy effect.
\end{titlepage}
\newpage

\renewcommand{\thefootnote}{\arabic{footnote}}
\setcounter{footnote}{0}
\sect{Introduction}
In recent years there have been much progress about 
 non-perturbative aspect in superstring theories. 
One of the non-trivial aspects of the superstring compactification is that the type-II stings with Calabi-Yau compactifications are connected by conifold transitions\cite{Reid,CDLS}. Physically these are interpreted as the transitions through black hole condensation\cite{St1,GMS,CGGK,Greene}. Later, it was shown that some of these transition can be understand as ordinary higgs transition. These are discovered as a consequence of heterotic-type-II duality\cite{KV} where K3 fibered Calabi-Yau manifolds are shown to be a potential candidate for the dual of heterotic string compactification on $K3\times T^2$\cite{KLM,AL,A}.  
There has been many non-trivial check of this duality at one-loop order in 
refs.\cite{CDFLL,KLT,HaM,Koko,CCLM,CCL,HeM,Kawai,C,M}.  
The non-perturbative check has been given by showing the relation to 
Seiberg-Witten theory\cite{KKLMV,KLMVW,Lerche,Kle}.
Afterward, the generic strategy to read off the gauge groups from toric 
diagrams has been discussed\cite{CPR}. 
  
As a practical analysis of the prepotential of the models, several methods has been studied in large moduli limit where we can argue them from mirror conjecture originated in ref.\cite{CdGP}. Convincing the mirror hypothesis with a consideration of  monodromies, a systematic analysis has been established\cite{CdFKM,CFKM,HKTY1,HKTY2}. On the other hand as for the prepotentials around conifold locus, there seems to be no systematic approach for the 
evaluation. However, in the case of enhanced gauge symmetry, an    
analysis has been given in ref.\cite{KKLMV} where they solve the solutions of Picard-Fuchs equation around the points of enhanced gauge symmetries and the basis are chosen in such a way that the basis recovers the one in Seiberg-Witten basis in the limit $\alpha' \rightarrow 0$. In other words,  the duality 
conjecture can be employed for the determination of prepotentials 
around the conifold locus, just 
like mirror conjecture have been 
in large moduli limit. As for the solutions of 
Picard-Fuchs equation, it is known that these are solved in the form of hypergeometric functions\cite{B,BS,HKTY1,HKTY2} in the large moduli region. However, these 
preferable characterization seems to be lost around conifold locus although it may persists in some special cases\cite{Suzuki1,Suzuki2}. Off course, it is not an essential problem of the system whether solutions can be solved in 
compact expression. 
The main problem of the approach is how to determine periods from 
the solutions of Picard-Fuchs equation. The usual approach for this is 
the consideration 
of monodoromy\cite{CdGP,CdFKM,CFKM}.  However, the analysis for the generic situations, we may need the consideration of monodromies or the analytic continuation from the large moduli limit, which is expected to be more difficult when we have many moduli parameters.  Therefore, it is favorable to have a new tool to determine periods at various region of moduli parameters. 

In this article, we will construct bi-linear symplectic form on the solution space of Picard-Fuchs equations. Namely, we are going to consider a symplectic map from the solutions of the equations to c-numbers.  Because it is manifestly invariant under monodromy transformations, we can determine the candidates of the periods up to the overall normalization which may also be determined by a simple consideration of monodromy.  These bi-linear form will turn out to be related to the existence of flat coordinates, which will be explicitly demonstrated for one and two moduli models. In two moduli models which are shown to have enhanced gauge symmetries, we will show that the choice of periods determined by duality are consistent with the analysis of bi-linear form of the models which implies the integrability of the prepotentials. We will also list the explicit form of the prepotential around the point of the enhanced gauge symmetries from which we may argue how the stringy correcti!
ons 
enter for $SU(2)$ Seiberg-Witten theories\cite{SW}. 
 
This article will be organized as follows. 

In section 2, 
we will show how to obtain symplectic form on the solution space of 
Picard-Fuchs equation by 
using the example of $1-$moduli models and the $2-$moduli 
models of $K3$ fibered Calabi-Yau three-fold. 
In section 3, we will demonstrate that the existence of flat 
coordinates are related directly 
to the symplectic form in one and two moduli models. 
In section 4,  we will consistently specify 
the dual pair of fields around the conifold point in one and two moduli 
models by using bi-linear form. We will also list the prepotential around the 
point of enhanced gauge symmetries in two moduli models.
Last section is devoted to the conclusion and some discussion.
\section{Intersection form on the solution space of the
Picard-Fuchs equations}
In this section, we investigate how to determine 
the symplectic form of the solution space of the Picard-Fuchs equations 
 associated to type II string theory. The construction of the 
symplectic basis defined in Gauss-Manin system is 
apparent for integrability and invariance of monodromy transformation. 
Thus we discuss first in this system though 
this system is equivalent to the Picard-Fuchs system. 

Consider mirror manifold $\bar M$ of Calabi-Yau 3-fold $M$. 
Canonical homology basis of $H_3(\bar M,Z)$ is $\{A^a,B_a\}\ (a=0,
1,\cdots,h^{2,1})$ and dual cohomology basis $\{\alpha_a,\beta^a\}$ of
 $H^3(\bar M,Z)$  are given by 
\bea
\int_{A^a}\alpha_b=\delta^a_b,\ 
\int_{B_a}\beta^b=\delta_a^b.\ \ (a,b=0,1,\cdots,h^{2,1})
\ena
The period integral $f_i(\mu_a)$ is defined by
\bea
f_i(\mu_a)=\int_{\gamma_i}\Omega(\mu_a),
\ena
where $\Omega$ is a holomorphic 3-form, $\gamma_i$ is a homology cycle
 in $H_3(\bar M,Z)$ and $\mu_a\ (a=1, \cdots, h^{2,1})$ 
are moduli parameters of complex structure deformation. 
 
Now consider Gauss-Manin system\cite{Griffiths};
\bea
{\pa \over \pa \mu_a}\Pi_i=M_a\Pi_i,\ \ (i=1,\cdots,2h_{2,1}+2)
\ena
where $\Pi_i={}^t(f_i,\pa_{\mu_a}f_i,\cdots )$ is 
$(2 h^{2,1}+2)$-dimensional vector and  $M_a$ 
is a $(2h^{2,1}+2)\times (2h^{2,1}+2)
$ matrix. This system reduces to the Picard-Fuchs system satisfied by $f_i$. 
Associated to Gauss-Manin system, 
we define the symplectic form $C_{ij}$, as  
\bea
<f_i,f_j>=\Pi_i^{T}A\Pi_j=C_{ij},\ \ A^{T}=-A,
\ena
where second condition insures anti-symmetric form  $C_{ij}=-C_{ji}$. 
On requiring $C_{ij}$ to be independent from moduli parameter, 
we have to 
solve the equation ${\pa\over \pa \mu_a}<f_i,f_j>=0$, i.e., 
\bea
{\pa\over \pa \mu_a}A+M_a^{T}A+AM_a=0.
\ena
Once we find $A$, we get the intersection form $C_{ij}$ which defines 
 moduli-independent norm on solution space. 
Under these conditions, it is easy to check the integrability 
of this system by using (2.1) and (2.3).
 Furthermore   
 this form is manifestly invariant under monodromy transformation. Therefore 
the intersection 
form have the information enough to construct the period vector 
canonically. 
Namely, if we label $f_i$ as $f_a^A,f_a^B$ obeying following 
combinations
\bea
<f_a^A,f_b^A>=<f_a^B,f_b^B>=0,\ \ <f_a^A,f_b^B>=\delta_{a,b},
\ena
and set $f_a^A$ to be a solution for $A^a$ cycle, then  
 $f_a^B$ corresponds to a solution along $B_a$ cycle naturally. 
For the canonical period vector $\{z_a,{\cal F}^a\}$, 
 we identify $f_a^A=z_a,\ f_b^B={\cal F}^b$ from this property 
up to normalization which undertaken by $Sp(2h^{2,1}+2,R)$ transformation, 
 where $\{z_a,{\cal F}^a\}$  are defined by
\bea
z_a&=&\int_{A^a}\Omega,\ \ 
{\cal F}^a=\int_{B^a}\Omega.\\
<z_{a},{\cal{F}}^b>&=&\delta_a^b,\ <z_a,z_b>=<{\cal F}^a,{\cal F}^b>=0, 
\ena

In this article,  
we intend to calculate prepotential of 
type II string theory by fixing period vectors from 
 the Picard-Fuchs equation directly. 
So let us to discuss how to do this in the Picard-Fuchs system. 
In order to clarify the symplectic structure on Picard-Fuchs equations, we will use the following notation on the space of differential operators.
 Introducing bi-linear operator
\bea
B\wedge D(f_1,f_2)={1 \over 2}(Bf_1\,Df_2-Df_1\,Bf_2),
\ena
where $B,D$ are any differential operators with respect to moduli 
parameters $\mu_a$, 
we will write the symplectic form in the form.
\bea
C_{ij}=C(f_i,f_j)=\sum_{k,k'}A_{(k,k')}D^{(k)}\wedge D^{(k')}\,(f_i,f_j),
\ena
where $A^{(k,k')}$ are functions of $\mu_a$ and $ D^{(k)}$ are the $k$-th 
order differential operators.  
In order to obtain $C_{ij}$ we have to find the combination 
of operator $ D^{(k)}\wedge D^{(k')}$ in such a way that we have 
\bea
{\pa_{\mu_a}}C=0,
\ena
using the ring of differential operators given by Picard-Fuchs equation. 
If we can do this, next 
we solve this equation for $A_{(k,k')}$. This is an alternative way 
to find $C_{ij}$ in the Picard-Fuchs system. Off course there is no 
guarantee that we will be able to  have simple solutions for the coefficients 
$A_{(k,k')}$. However, as an explicit evaluations of the solutions for a 
few moduli models, we will obtain rather simple form of the symplectic 
forms, which will be considered in the next section.    
\subsection{$1- $moduli models}
Before turning to the main subject of this article, 
we are going to perform how to obtain $C_{ij}$ with the 
 models of Calabi-Yau three-fold constructed from 
the hypersurfaces in the 
toric varieties and Grassmannian with $1- $moduli as the basic examples. 

For the models constructed by hypersurfaces in toric variety\cite{B}, 
 Picard-Fuchs equation in 
the large complex structure limit can be written as
\bea D\,f=
\left[\th_x^4-x(\th_x+\la_1)(\th_x+1-\la_1)(\th_x+\la_2)(\th_x+1-\la_2)
\right]f=0,
\ena
where $\th_x=x{\pa\over \pa x}$, $x$ is a variable made 
of the moduli of these manifolds being 
$x\sim 0$ in this region, and $\la_i\ (i=1,2)$ is 
a rational number associated to each manifold\cite{KT,LT}. For example, the  
famous quintic model $P^4[5]$ corresponds to a model with 
$\la_1={1\over 5}, \la_2={2\over 5}$. 

Let us consider the symplectic form of the solution space of above 
Picard-Fuchs equation. 
Since the Picard-Fuchs equation is of forth order, 
we propose the intersection form $C$ 
 up to  third order of $\th_x$ in the form:
\bea
C=A_1\,1\wedge \th_x^3+A_2\,\th_x\wedge \th_x^2+
A_3\,1\wedge\th_x^2+A_4\,1\wedge \th_x,
\ena
where coefficients $A_i\ (i=1,\cdots 4)$ are some functions
 of $x$ to be determined. We impose the condition 
\bea
\th_x\,C=0,
\ena
in the ring $D=0$, so that  the intersection of the solution space $<f_1,f_2>=C(f_1,f_2)$ is a constant for any pair of solutions of Picard-Fuchs equation. Applying $\th_x$ to (2.9) and 
using (2.7) to descend the order of $\th_x^4$, we derive  
equations which have to be obeyed by $A_i$. The solution of the equations turns out to be very simple. We find  the bilinear form is given by
\bea
C&=&(1-x)\left\{
1\wedge \th_x^3 -\th_x\wedge \th_x^2\right\}
\nonumber \\
&-&x\left\{1\wedge \th_x^2(f_1,f_2)+
[\la_1(1-\la_1)+\la_2(1-\la_2)]\,1\wedge \th_x\right\}.
\ena 
It seems interesting that the Wronskian of the Picard-Fuchs equation can be obtained by simple manipulation of $C$ as $C\wedge C \sim (1-x) 1\wedge \th_x \wedge \th_x^2 \wedge \th_x^3$. In other words, $C$ can be understood as a square root of Wronskian.

Let us see how this works when we 
construct the basis. In large moduli limit $x\sim 0$, 
 a power solution of the Picard-Fuchs equation (2.6) can be written as
\bea
W_0=\sum_{n=0}^{\infty}{(\la_1)_n(1-\la_1)_n(\la_2)_n(1-\la_2)_n
\over \Gamma(n+1)^4}x^n,
\ena
where $(a)_n=\Gamma(a+n)/\Gamma(a)$. 
Other three solutions, $W_1,W_2, 
W_3$, can be given by Frobenius method; define 
the operator $D_{\rho_i}$ applying to generic power solution
 $\sum_{n_i}c(n_i)x^{n_i}$,  
\bea
D_{\rho_i}c(n_i)\,x^{n_i}=\left.{\pa_{\rho_i}
c(n_i+\rho_i) \over 2\pi i}\,x^{n_i+\rho_i}\right|_{\rho_i=0}.
\ena
other three solutions can be written by
\bea
W_1=D_{\rho_x}W_0,\ \ 
W_2=(D_{\rho_x})^2W_0,\ \  
W_3=(D_{\rho_x})^3W_0.
\ena
In the leading order, these solution symplectic form can be written as
\bea
& W_0 &\sim 1,\ \ W_1\sim {\log x \over 2 \pi i },\ \ 
W_2\sim { (\log x)^2 \over (2 \pi i)^2 },\ \ W_3\sim { 
(\log x)^3 \over (2 \pi i)^3 },\nonumber\\
& C &\sim 1\wedge \th_x^3-\th_x\wedge \th_x^2,
\ena
Substituting these solutions directly to $C$ in various combination, 
it is easy to see that only $C(W_0,W_3)$ 
and $C(W_1,W_2)$ are not zero, and 
all other combination vanish in the lowest order. 
 Even including higher order terms, the intersection 
of these solutions are not changed. 
Therefore it is natural to conclude that the symplectic bases taken by 
these solutions are 
\bea
z^0\sim W_0,\ z^1\sim W_1,\ 
{\cal F}_0\sim W_3,\ {\cal F}_1\sim W_2,
\ena
where we impose the asymptotic behavior of $t=z^1/z^0$ is 
$t\sim \log x/(2\pi i)$. This means that 
if we fix the component of $\alpha$ cycle, the intersection 
form selects 
proper $\beta$ cycle automatically. The intersection form by itself 
 determines only whether the intersection of 
 two solutions is zero or not, and 
 this does not fix the normalization for the symplectic basis. 
In order to fix the normalization, we just need the classical yukawa coupling.
 Then, the prepotential is given by definition as 
\bea
{\cal F}={1\over 2}{z^a{\cal F}_a \over (z^0)^2},
\ena
and inverting $t$ as $x=x(t)$, and substituting this to 
$\cal F$, the prepotential in this region is written 
 as the function of $t$, which is equivalent to the 
result of \cite{HKTY1,HKTY2,H}.

In this way, the normalized bi-linear form can be obtained in the form;
\bea
C&=&{(2\pi i)^3\over \kappa_0}\left[(1-x)\left\{
1\wedge \th_x^3 -\th_x\wedge \th_x^2\right\}\right.
\nonumber \\
&-&x\left.\left\{1\wedge \th_x^2(f_1,f_2)+
[\la_1(1-\la_1)+\la_2(1-\la_2)]\,1\wedge \th_x\right\}\right],
\ena 
where $\kappa_0$ is the classical yukawa coupling 
and $\kappa_0=16\sin^2\la_1\pi\sin^2\la_2\pi$ for 
simple series of models \cite{KT,LT}. This normalized 
form can be considered as the intersection form on periods.

Next we extend the discussion to models constructed from the 
  hypersurfaces in the Grassmannian considered recently in refs.\cite{BFKS}. 
A fundamental period 
in the large moduli limit  can be obtained as diagonal degeneration of the periods of complete intersection of projective space\cite{BFKS}. 
For the examples of one moduli models listed in ref.\cite{BFKS}, 
Picard-Fuchs operator of order 4 can be written in the following form as
\bea
\th_x^4=
\sum_{i=1}^r a_ix^i(\th_x^2+i\,\th_x+\alpha_i)
(\th_x^2+i\,\th_x+\beta_i),
\ena
where $a_i,\alpha_i,\beta_i$ are rational number 
associated to each case, and this reduces to the toric variety 
with $r=1$ complete intersection with 
$a_1=1,\ \alpha_1=\la_1(1-\la_1),\ 
\beta=\la_2(1-\la_2)$.  Along the same line 
in the case of toric variety, we can give 
the bi-linear form 
\bea
C=(1-\sum_{i=1}^r a_ix^i)\left\{
\th_x\wedge \th_x^2-1\wedge \th_x^3\right\}+
\sum_{i=1}^ri\,a_ix^i\,1\wedge \th_x^2+
\sum_{i=1}^r(\alpha_i+\beta_i)x^i\,1\wedge \th_x.
\ena
 It is straightforward to specify periods even in these models. 
\subsection{$2-$parameter models with $K3$-fibration}
In this subsection we deal with a series 
of $2-$moduli models of type II string theory compactified 
on Calabi-Yau three-fold constructed by 
K3 fibrations.  
This series consist of four models \cite{KLM} denoted as
 $P^{1,1,2,2,6}[12],\ 
P^{1,1,2,2,2}[8],$ $P^{1,1,2,2,2,2}[6,4],\ 
P^{1,1,2,2,2,2,2}[4,4,4]$. In these series, 
moduli parameters $\psi $ and $\phi$ are combined to 
$x$ and $y$ as
\bea
x={\phi\over \psi^{1\over \la}},\ y={1\over \phi^2},
\ena
where $\la$ is ${1\over 6},\ {1\over 4},\ {1\over 3},\ {1\over 2}$  
for $P^{1,1,2,2,6}[12],\ 
P^{1,1,2,2,2}[8],$ $P^{1,1,2,2,2,2}[6,4],\ 
P^{1,1,2,2,2,2,2}[4,4,4]$, \\
respectively\cite{Suzuki2}. 
Their Picard-Fuchs equations can be 
given by following set of operators
\bea
D_1&=&\th_x^2(\th_x-2\th_y)-x(\th_x+\la)(\th_x+{1\over 2})(\th_x+
1-\la),\\
D_2&=&\th_y^2-{1\over 4}y(2\th_y-\th_x+1)(2\th_y-\th_x),
\ena
or equivalently,
\bea
D_1&=&(1-x)\th_x^3-2\th_x^2\th_y-{3x\over 2}\th_x^2-[
{1\over 2}+\la(1-\la)]\,x\th_x-{\la(1-\la)\over 2}\,x,\\
D_2&=&(1-y)\th_y^2+y\th_x\th_y-{y\over 4}\th_x^2-{y\over 2}\th_y
+{y\over 2}\th_x, 
\ena
Now let us consider the combination of 
bilinear operators to construct the intersection form 
 close in the order of $\th$ as fewer as possible. 
 The first order is trivial; $1\wedge \th_x$ and 
$1\wedge\th_y$. In second order there are 
four possibility; $1\wedge \th_x^2, \ 1\wedge \th_y^2,\ 
1\wedge \th_x\th_y$, $\th_y\wedge \th_x$. However 
one of these are not independent and we eliminate 
$1\wedge \th_y^2$ by using (2.18). 
 Similarly in third order there are ten possibility, however, 
four terms which contain $\th_y^2$, such as 
$1\wedge \th_x\th_y^2,\ \th_x\wedge\th_y^2,\  \th_y\wedge\th_y^2$
 and $1\wedge\th_y^3$, are descended by (2.18), and 
a term which contains $\th_y^3$ are eliminated by (2.17). 
Therefore in the third order we have five independent operator. 
Since the Picard-Fuchs operators are at most of order three in this series, 
we expect the intersection form close up to third order to be 
\bea
C&=&A_1\,1\wedge\th_x^2\th_y+A_2\,\th_y\wedge \th_x^2+
A_3\th_x\wedge\th_x\th_y+ A_4\, \th_x\wedge \th_x^2+A_5\, \th_y\wedge
 \th_x\th_y\nonumber \\
& &+
A_6\,1\wedge\th_x^2+A_7\,1\wedge\th_x\th_y+A_8\,\th_y\wedge\th_x+
A_9\,1\wedge\th_x+A_{10}\,1\wedge \th_y.
\ena
In order to determine coefficients $A_i\ (i=1,\cdots 10)$, 
we impose following conditions
\bea
\th_xC=\th_yC=0,
\ena
in the ring of differential operators $D_1 =D_2 =0$.

These will results in the first order differential equations for the coefficients. Although the equations seems to be messy, 
due to these relations, 
we can solve $A_i$ in a simple form to find
\bea
C&=&{\Delta\over 2(1-x)}\,1\wedge\th_x^2\th_y
-{(1-x)(1-y)\over 2}\,\th_y\wedge \th_x^2
-{(1-x)(1-y)\over 2}\,\th_x\wedge\th_x\th_y\nonumber \\
& &+{(2x-1)y\over 4}\, \th_x\wedge \th_x^2+
(1-y)\, \th_y\wedge
 \th_x\th_y -
{yx(2x+1)\over 8(1-x)}
\,1\wedge\th_x^2 \nonumber \\
 & &-
{x(1-y)\over 4}\,1\wedge\th_x\th_y+
{x(1-y)\over 4}\,\th_y\wedge\th_x\\
 & &+
{xy(x^2+(-1+2\la(1-\la))x-
2\la(1-\la))\over 8(1-x)}\,1\wedge\th_x-
{x(1-y)\,\la(1-\la)\over 2}\,1\wedge \th_y,\nonumber 
\ena
where $\Delta=(1-x)^2-yx^2$ is the discriminunt of the theory. 

Let us see how above bi-linear form recovers canonical symplectic base  
 which have been obtained previously. 
In the region $x\sim 0,\ y\sim 0$, power solution can be 
given by
\bea
W_0=\sum_{n,m}{(\la)_m({1\over 2})_m(1-\la)_m\over 
\Gamma(m-2n+1)\,(m!)^2(n!)^2}x^m\left({y\over 4}\right)^m,
\ena
and other five solution are obtained by using Frobenius methods 
as before. As long as 
considering whether the intersection is zero or not, 
the normalization of the solution does not matter. 
 Thus from the incidial equation we take logarithmic solutions as 
\bea
W_1&=&D_{\rho_x}W_0,\ \ \ 
W_2=D_{\rho_y}W_0,\\
W_3&=&D_{\rho_x}D_{\rho_x}W_0,\ \ 
W_4=(D_{\rho_x}D_{\rho_y}+D_{\rho_x}^2) W_0,\\
W_5&=&({2\over 3}D_{\rho_x}^3+D_{\rho_x}^2D_{\rho_y})W_0
\ena
Furthermore, it is enough to calculate in the leading order when we 
construct the basis of period vectors. 
 All solutions in 
the leading order can be written 
\bea
&W_0&\sim 1,\ W_1\sim \log x,\ W_2\sim \log y,\\
&W_3&\sim 
(\log x)^2,\ \ W_4\sim (\log x+\log y)(\log x),\\
&W_5&\sim 
{2\over 3}(\log x)^3+(\log x)^2\log y.
\ena
Also we can write the intersection form effectively 
in the leading order in the following 
form
\bea
C\sim {1\over 2}1\wedge \th_x^2\th_y-
{1\over 2}\th_y\wedge \th_x^2-{1\over 2}
\th_x\wedge \th_x\th_y.
\ena
Each constant can be calculated directly as
\bea
C(W_0,W_5)=1,\ \ C(W_1,W_4)=-1,\ \ C(W_2,W_3)=-1 
\ena
and the other combinations  vanish. 
Assuming that $t_{1}\sim z^1/z^0\sim \log x/(2\pi i) $ and 
$t_2\sim z^2/z^0\sim \log y/(2\pi i)$, we conclude to 
take following combination as the symplectic base, 
\bea
z^0\sim W_0,\ \ z^1\sim W_1,\ \ z^2\sim W_2,\\
{\cal F}_0\sim W_5,\ \ {\cal F}_1\sim W_4,\ \ {\cal F}_2\sim W_3,
\ena
Thus we can recover the result obtained in refs.\cite{CdFKM,HKTY1,HKTY2}. 
 In the connection to the normalization, 
the prepotential and Yukawa coupling of this series of models can 
be given by 
\bea
{\cal F}&=&-{1\over 3!}K_{ijk}t^it^jt^k+\cdots,\\
K_{111}&=&2K_{211}=8\sin^2 \la\pi.
\ena
All solutions are written with 
the normalization of the convention of \cite{HKTY1} by 
using Yukawa coupling in appendix B. However, as was mentioned, 
 we now concentrate on deriving the prepotential 
in one region, it is not 
necessary to fix the over all constant of the intersection form. 
All we need is the combination of the solution with non-zero 
intersection, and the relative normalization of Yukawa 
coupling of the tree level part of 
the prepotential, and also proper asymptotic behavior of $\alpha$ cycle. 

\sect{Flat coordinate condition and bi-linear intersection form}
This section is motivated by the observation in terms of 
the flat coordinate given in 
refs.\cite{HL,No}. 
In string theory, 
you can calculate exact Yukawa coupling 
constant including instanton effect in the large 
radius limit due to the mirror symmetry \cite{CdGP,HKTY1,HKTY2}. Along 
this calculation, flat coordinates 
 are regarded as 
the special affine coordinates made of $\alpha$ cycles \cite{St2}; these 
 are used to fix the combination of solution 
of Picard-Fuchs equation, 
so as to establish the property of maximal unipotent monodromy 
\cite{CdGP,HKTY1,HKTY2}.  
Recently interesting connection between periods 
and flat coordinate are worked out in \cite{HL,No}. 
 In these analysis, 
 mirror symmetry can be represented 
as isomorphic pair of quantum cohomology ring of 
different manifold. By changing 
the basis of Jacobian ring,  
Gauss-Manin system can be written to 
 a form with flat connection in terms of special coordinates; 
 flat coordinates. 
 Consequently the specification of the periods from these conditions 
 completely match the results with regard to the 
maximal unipotency of monodromy \cite{CdGP,HKTY1}. Since this monodromy 
property insures symplectic structure of periods, 
above analysis have to do with our 
procedure of specification of periods 
 based on bi-linear form in the large radius 
limit.  Our aim in this section is to show that bi-linear form we have 
constructed are directly related to the condition of flat coordinates. 

\subsection{$1-$moduli models}

In this subsection we consider Gauss-Manin system of 
the Calabi-Yau space constructed from toric variety of 
 $1-$moduli models discussed in Sect. 2.1. 
Following refs.\cite{HL,No}, we consider 
a set of first order differential equations: 
\bea
\theta_x w=G\,w,\label{eq:gm1-1}
\ena
where $w$ is a vector which consists of period integrals. 
 We prefer to use $\theta_x$ rather ${\pa\over \pa x}$ to have to do 
with the result of last section, and choose 
basis to take $w={}^t(f_i, \theta_xf_i,\theta_x^2f_i,\theta_x^3f_i)$ and 
$f_i$ is given by (2.2) in the case of $1-$moduli models.  
The Picard-Fuchs operator can be rewritten 
in the ring $D=0$ by making use of (2.12) as 
\bea
\theta_x^4=a_4\theta_x^3+a_3\theta_x^2+a_2\theta_x+a_1,
\ena
where 
\bea
a_1&=&{\la_1(1-\la_1)\la_2(1-\la_2)\over 1-x}x,\ 
a_2={\la_1(1-\la_1)+\la_2(1-\la_2)\over 1-x}x,\nonumber \\
a_3&=&{\la_1(1-\la_1)+\la_2(1-\la_2)\over 1-x}x,\ 
a_4={2x\over 1-x}.
\ena
In order for the Gauss-Manin system (\ref{eq:gm1-1}) to 
reduce the Picard-Fuchs equation (2.12), we take 
the matrix $G$ as
\bea
G=\left(\begin{array}{cccc}
0&1&0&0\\
0&0&1&0\\
0&0&0&1\\
a_1&a_2&a_3&a_4
\end{array}\right)
\ena

Let us introduce the flat coordinates $t=t(x)$ and 
new periods $v(t)$,  
to rewrite the Gauss-Manin system for 
$v$. Flat coordinate conditions can be represented as 
 couplings of new basis given by degree preserving 
transformation of deformed Jacobian ring, which 
is reduced to the Gauss-Manin system for 
$v$ in the following form:
\bea
\theta_t v(t)=R\,v(t)= 
\left(\begin{array}{cccc}
0&1&0&0\\
0&0&K_{ttt}(t)&0\\
0&0&0&1\\
0&0&0&0
\end{array}\right)v(t).
\ena
where $K_{ttt}(t)$ is Yukawa coupling in this coordinate. 
New period $v$ is related to original one by
\bea
w=M(x)\,v,
\ena
and it is known that 
transfer matrix $M(x)$ can be taken as a lower-triangle matrix
\bea
M=\left(\begin{array}{cccc}
r_{11}&0&0&0\\
r_{21}&r_{22}&0&0\\
r_{31}&r_{32}&r_{33}&0\\
r_{41}&r_{42}&r_{43}&r_{44}
\end{array}\right).
\ena
By substituting (3.5) to 
 (3.1) and rewriting to the Gauss-Manin system for $v$, we derive 
the condition to determine the transfer matrix $M$,
\bea
A:=GM-\theta_x \,M-\theta_xt\,M\,R=0.
\ena
From the analysis about conditions  $A_{i1}=0\ (i=1,\cdots , 4)$,  
 $r_{11}$ must be the solution of the Picard-Fuchs equation, 
so we set $r_{11}=g$. Also after some algebra, 
we see that once we know concrete form of  
$r_{44}$, every $r_{ij}$  and 
Yukawa coupling $K_{ttt}$ can be represented by using
$\theta_x t$ and $r_{11}$. To accomplish this, using components $A_{23}, 
A_{33}, A_{34}, A_{44}$ we derive 
the differential equation satisfied by $r_{44}$, so that 
we can determine the form of $r_{44}$:
\bea
r_{44}={1\over (1-x)g}.
\ena
Thus substituting this to all components but $A_{43}$, we 
obtain the complete form of the transfer matrix $M$. 

Now let us show how flat coordinate condition 
relate to the bi-linear intersection form. The only 
component which have not been used to obtain $r_{ij}$ is 
$A_{43}=0$, or 
\bea
&2&\theta_xg\theta_x^2t+4\theta_xt\theta_x^2g
+g\theta_x^3t-2{(\theta_x g)^2\over g}\theta_xt\\
&-&{x\over 1-x}\left\{g\theta_x^2t+(\la_1(1-\la_1)+\la_2(1-\la_2))g
\theta_xt-2\theta_xt\theta_xg\right\}=0.\nonumber
\ena
As was claimed in ref.\cite{No}, the role 
of this condition is to determine 
the flat coordinate $t(x)$. Here we assume 
 $t={f/ g}$, where  $f$ is a solution of the Picard-Fuchs equation. 
Although $t$ is not necessary a ratio of periods, this assumption will
make clear the relation between choice of the flat coordinate 
and specification of periods. Indeed, the condition $A_{43}=0$ can be 
rewritten as the bi-linear intersection form for $f$ and $g$ itself!
\bea
C(g,f)=0.
\ena
This means that in the $1-$moduli models, 
to search for the flat coordinate under the above assumption is 
equivalent to determine the periods as a combination 
of the solutions of the Picard-Fuchs equation. Thus in the case of analyzing 
mirror symmetry, we may always take the flat coordinate 
in the large radius limit 
as the ratio of independent $\alpha$ cycles. 
 The bi-linear intersection form is a 
key to explain this consequence explicitly. 

In the $2-$moduli models, 
the situation become slightly different from above in that 
the flat coordinate condition does not necessary coincide 
the condition of vanishing intersection of each period, 
as we will see in the next subsection. 

\subsection{$2-$moduli models}

In this subsection we investigate how 
the flat coordinate condition relate to 
the bi-linear intersection form in $2-$moduli models discussed in Sect.2.2.

Hereafter we set the basis of the ring $D_1=D_2=0$ in
 $2-$moduli models of (2.26) and (2.27) as 
$\{L^{(1)},L^{(2)},L^{(3)},L^{(4)},L^{(5)},L^{(6)}\}=
\{1,\theta_x,\theta_y,\theta_x^2,\theta_x\theta_y,\theta_x^2\theta_y\}$. 
In this case, the Gauss-Manin system which 
recovers (2.26) and (2.27) can be written by
\bea
\theta_x \,w=G\,w,\ \ 
\theta_y\, w=H\,w,
\ena
where $w={}^t(f_i,L^{(2)}f_i,
L^{(3)}f_i, L^{(4)}f_i,
L^{(5)} f_i,L^{(6)}f_i)$, and 
$G$ and $M$ are given by
\bea
G=\left(\begin{array}{cccccc}
0&1&0&0&0&0\\
0&0&0&1&0&0\\
0&0&0&0&1&0\\
a_1&a_2&a_3&a_4&a_5&a_6\\
0&0&0&0&0&1\\
a'_1&a'_2&a'_3&a'_4&a'_5&a'_6
\end{array}\right)
,\ \ H=\left(\begin{array}{cccccc}
0&0&1&0&0&0\\
0&0&0&0&1&0\\
b_1&b_2&b_3&b_4&b_5&b_6\\
0&0&0&0&0&1\\
b'_1&b'_2&b'_3&b'_4&b'_5&b'_6\\
b''_1&b''_2&b''_3&b''_4&b''_5&b''_6
\end{array}\right),
\ena
here we denote $a_i$ etc. as 
coefficients for following operators in the ring $D_1=D_2=0$
\bea
\theta_x^3&=&a_iL^{(i)},\ \ \theta_x^3\theta_y=a'_iL^{(i)},\\
\theta_y^2&=&b_iL^{(i)},\ \ \theta_x\theta_y^2=b'_iL^{(i)},\ \ 
\theta_x^2\theta_y^2=b''_iL^{(i)}, 
\ena
and all coefficients  $a_i,a'_1,b_i,b'_i,b''_i\ (i=1,\cdots , 6)$ 
are immediately derived from the ring $D_1=D_2=0$ after 
some algebra, though we shall not list them 
 here. By introducing flat coordinates 
$t=t(x,y), s=s(x,y)$, and new period $v$ relating to original one $w$ as
\bea
w&=&M\,v
\ena
we intend to rewrite the Gauss-Manin system for $v$ in the following form:
\bea
\theta_t\,v&=&R_t\,v,\ \ \ 
\theta_s\,v=R_s\,v,
\ena
where, as the reduction of the flat coordinate condition 
of of new basis of deformed Jacobian ring\cite{No}, 
matrix $R_t$ and $R_s$ are given by
\bea
R_t&=&
\left(\begin{array}{cccccc}
0&1&0&0&0&0\\
0&0&0&K_{tts}&K_{ttt}&0\\
0&0&0&K_{tss}&K_{tst}&0\\
0&0&0&0&0&0\\
0&0&0&0&0&1\\
0&0&0&0&0&0
\end{array}\right)
,\ \ 
R_s=
\left(\begin{array}{cccccc}
0&0&1&0&0&0\\
0&0&0&K_{sts}&K_{stt}&0\\
0&0&0&K_{sss}&K_{sst}&0\\
0&0&0&0&0&1\\
0&0&0&0&0&0\\
0&0&0&0&0&0
\end{array}\right). 
\ena
In $2-$moduli models, it is known 
that $M$ needs non-vanishing components additionally to be a 
 lower-triangle form as
\bea
M&=&\left(\begin{array}{cccccc}
r_{11}&0&0&0&0&0\\
r_{21}&r_{22}&r_{23}&0&0&0\\
r_{31}&r_{32}&r_{33}&0&0&0\\
r_{41}&r_{42}&r_{43}&r_{44}&r_{45}&0\\
r_{51}&r_{52}&r_{53}&r_{54}&r_{55}&0\\
r_{61}&r_{62}&r_{63}&r_{64}&r_{65}&r_{66}
\end{array}\right).
\ena
By substituting (3.15) to (3.11) and comparing to (3.16), 
the condition to determine $M(x,y)$
is derived:
\bea
A:=GM-\theta_x\,M-\theta_x t\,MR_t-\theta_xs\,MR_s=0,\\
B:=HM-\theta_y\,M-\theta_y t\,MR_t-\theta_ys\,MR_s=0.
\ena
After some algebra, we see immediately 
 $r_{11}$ must be the solution of the Picard-Fuchs equation 
(2.26) and (2.27), so we set $r_{11}=g$ as before. Also we see that, 
as was pointed out in 
the ref.\cite{No}, some components coincide as 
\bea
A_{i4}=A_{i5},\ \ B_{i4}=B_{i5}\  (i=4,5,6).
\ena
In order to know all $r_{ij}$, we must obtain $r_{66}$. 
To do this we concentrate on two conditions, $A_{44}=0, A_{54}=0$.  
Explicit calculation shows that 
there are identical terms in both conditions up to factors. Consequently  
$\theta_yt A_{44}-\theta_xtA_{54}=0$ will be 
the differential equation satisfied by $r_{66}$:
\bea
& &(\theta_xt)^2\left\{
-b'_6\theta_x-\theta_y-b'_6{\theta_xg\over g}-{\theta_yg\over g}
+b''_6-(\theta_xb'_6)\right\}r_{66}\nonumber \\
&+&\theta_xt\theta_yt\left\{
3\theta_x+a_6\theta_y+3{\theta_xg\over g}+a_6{\theta_yg\over g}
-a_4-a_6b''_6+a'_6\right\}r_{66}\\
&+&(\theta_yt)^2\left\{
-2a_6\theta_x-2a_6{\theta_xg\over g}+a_4a_6+a_6a'_6-(\theta_xa_6)\right\}
r_{66}=0.\nonumber
\ena
By solving this equation, we obtain $r_{66}$:
\bea
r_{66}={1-x\over \Delta\,g},
\ena
where $\Delta=(1-x)^2-x^2y$ is the discriminunt. By substituting 
this to suitable components of the condition (3.19) and (3.20), 
we can know exact form of $M$ and Yukawa couplings, though 
 we will not list these results here. We will rather 
turn to solve the flat coordinate
 condition. There still remain following components which 
we have not use until now 
\bea
A_{44}=0,\ A_{54}=0,\ A_{64}=0,\nonumber\\
B_{44}=0,\ B_{54}=0,\ B_{64}=0. 
\ena
These should be used to determine the flat coordinates $t(x,y), s(x,y)$. 
 How many conditions are independent among them? 
As a matter of fact, careful calculations reveal that 
four components $A_{44}, A_{54}, B_{44}, B_{54}$ mean 
same condition up to factor: 
\bea
Q:&=&(\theta_xt\theta_ys-\theta_xs\theta_yt)\left(-{x\over 2(1-x)}+
{\theta_x g\over g}-a_6{\theta_y g\over g}\right)\nonumber \\
& &-\theta_x^2t(b'_6\theta_xs-\theta_ys)+\theta_x^2s(
b'_6\theta_xt-\theta_xt) \label{eq:flat2}\\
& &+\theta_x\theta_yt(\theta_xs-a_6\theta_ys)-
\theta_x\theta_ys(\theta_xt-a_6\theta_yt)=0.\nonumber  
\ena
Furthermore, after complicated manipulations, 
it turns out that rest components $A_{64}$ and $B_{64}$ are 
represented by using derivative and multiples of $Q$ as
\bea
A_{64}&=&-{(b'_6\theta_xt-\theta_yt)\over \Delta}\theta_x\,Q+
{(\theta_xt-a_6\theta_yt)\over \Delta}\theta_y\,Q+\{\cdots\}Q
=0,\\
B_{64}&=&-{\{(b_4+b_5b'_6)\theta_xt-b'_6\theta_yt\}\over \Delta}\theta_x
\,Q+{(b'_6\theta_xt-\theta_yt)\over \Delta}\theta_y\,Q+\{\cdots\}Q=0.
\ena
where we denote negligible multiples of $Q$ as dots in the brace. 
Therefore essentially flat coordinate condition is just (\ref{eq:flat2}). 
The question we have to ask now is how (\ref{eq:flat2}) relate to 
the bi-linear intersection form. Here 
 we assume the flat coordinates are the ratio 
of the Picard-Fuchs equation as
\bea
t={f_1\over g},\ s={f_2\over g},
\ena
where $f_1, f_2$ and $g$ are independent solutions to each other. 
Under this assumption, resulting expression of (\ref{eq:flat2}) 
will be completely 
anti-symmetric in terms of $g,f_1,f_2$, and 
the condition (\ref{eq:flat2}) can be rewritten by using 
the bi-linear intersection form (2.32) as
\bea
g\,C(f_1,f_2)+f_1\,C(f_2,g)+f_2(g,f_1)=0.
\ena
Thus we conclude that, contract to $1-$moduli models, a set of vanishing 
conditions of mutual intersection of periods is a mere 
sufficient condition of the flat coordinate condition. 
 Consequently, the solution space of the flat coordinate is 
larger than that of symplectic periods even 
under this assumption. So if you desire, 
you may find the solutions which 
satisfy the condition (\ref{eq:flat2}) although their mutual intersection 
does not vanish. Anyway, (\ref{eq:flat2}) represents manifestly 
the reason why the usual choice of the 
flat coordinate, which is a ratio of 
independent periods, work well.

\sect{Determination of prepotential around the conifold locus}
\vspace{-1.5cm}
\subsection{$1-$moduli models around the conifold
 locus}
In this subsection  we turn to evaluate the value of symplectic form 
around the conifold point $x=1$ in one moduli models. 
Changing the variable of (2.6) as $y=1-x$ and analyzing 
the incidial equation, we put four solutions around the 
conifold $Y_i\ (i=0,1,2,3)$ can be given in lowest order as
\bea
Y_0\sim 1,\ Y_1\sim y,\ 
Y_2\sim y\log y,\ Y_3\sim y^2. 
\ena
In this region, these solutions are obtained iteratively. 
The intersection of these solutions can be calculated 
directly in the leading order and 
we see that $C(Y_0,Y_3)$ and $C(Y_1,Y_2)$ are not zero 
and other combinations become zero. 
If we add higher order terms obtained by iterations, 
the intersection form (2.16) is not changed up to the higher orders as 
much as possible by mathematica. 
Taking conventional assumption for the asymptotic behavior for 
$\alpha$ cycle, 
it is natural to choose the symplectic base as 
\bea
\ti z^0\sim Y_0,\ \ \ti z^1\sim Y_1,\\
\ti {\cal F}_0\sim Y_3,\ \ \ti {\cal F}_1\sim Y_2, 
\ena
up to normalization. Around the conifold point, 
we cannot kill a degree of freedom such that we can  add 
 $\ti{\cal F}_0$ to $\ti z^0$ like $\ti z_0'=\ti z_0+\beta \ti{\cal F}_0$
, since this manipulation 
does not change the intersection of the basis. 
However, in other region $\ti {\cal F}$ may be  transformed 
by monodromy transformation, so from the 
global consistency of moduli space $\beta$ must be a integer number. 
 For maximal simplicity we set $\beta=0$. The normalization of the periods can be fixed by the behavior around conifold locus \cite{St1,GMS,Greene}: 
\bea
\ti{\cal F}_1 \sim {1\over 2\pi i} \ti{z}^1\log \ti{z}^1.
\ena
 Therefore, we can calculate the prepotential 
around conifold point once we can specify the periods of $\alpha$ cycle. 
This procedure 
will be described in detail in rather non-trivial 
example with $2-$moduli models in the following subsection. 
\subsection{prepotential around the conifold
 locus of $2-$moduli model}
In this section, we determine the period vector to give 
the exact prepotential of a series of $2-$moduli models around 
the conifold point. Conifold transition of the models are discussed in 
Ref.\cite{LS} and the enhanced gauge symmetries has been established in 
Refs.\cite{KKLMV,KLMVW}. 

First of all, we have to obtain the solution of the Picard-Fuchs 
equation around the conifold point. On Calabi-Yau three-fold obtained by
 K3 fibrations with 2-moduli, three kinds of the 
 power solution around the conifold 
point can be given in a systematic way\cite{Suzuki2}: 
\bea
Y_0&=&\sum_{n_1,n_2,m}{({\la\over 2})_{n_1}({1-\la\over 2})_{n_1}
({\la\over 2})_{n_2}({1-\la\over 2})_{n_2}\Gamma(n_1+n_2+1)
\over 
(n_1+n_2-2n)!({1\over 2})_{n_1}(n_1)!
({1\over 2})_{n_2}(n_2)!n!n!}x_2^{n_1+n_2}\left({x_1\over 4}\right)^m,\\
Y_1&=&\sum_{n_1,n_2,m}{({\la\over 2})_{n_1}({1-\la\over 2})_{n_1}
({1+\la\over 2})_{n_2}(1-{\la\over 2})_{n_2}\Gamma(n_1+n_2+{3\over 2})
\over 
\Gamma(n_1+n_2-2n+{3\over 2})({1\over 2})_{n_1}(n_1)!
({1\over 2})_{n_2}(n_2)!n!n!}x_2^{n_1+n_2+{1\over 2}}
\left({x_1\over 4}\right)^m,\\
Y_2&=&\sum_{n_1,n_2,m}{({\la\over 2})_{n_1}({1-\la\over 2})_{n_1}
({\la\over 2})_{n_1}({1-\la\over 2})_{n_1}\Gamma(n_1+n_2+2)
\over 
(n_1+n_2-2n+1)!({1\over 2})_{n_1}(n_1)!
({1\over 2})_{n_2}(n_2)!n!n!}x_2^{n_1+n_2+1}\left({x_1\over 4}\right)^m,
\ena
where $x_1=x^2y/(1-x)^2,\ x_2=1-x$. The other logarithmic solutions can be represented formally by barnes-type representations including well-poised series of type ${}_4F_3$\cite{Suzuki2}. However we could not find any good formula for the explicit evaluation of the representation.
 Fortunately, one solution $Y_4$ which is dual of 
$Y_1$, is given by ordinary manner as
\bea
Y_4=D_{\rho_y}Y_0.
\ena
In order to get remaining two solutions dual to $Y_1$ and $Y_3$, 
which contain $\log(x_1x_2^2)$ terms, 
 we have to solve the equation for the 
coefficient iteratively by setting the solutions in the form:
\bea
Y_3=\log(x_1x_2^2) Y_0+\sum_{n,m}p(n,m)x_1^nx_2^m,\\
Y_5=\log(x_1x_2^2) Y_2+\sum_{n,m}q(n,m)x_1^nx_2^m.
\ena
From the point of view of heterotic-type II string duality, 
 an enhancement of $SU(2)$ gauge symmetry occurs at the conifold point $x=1$ 
so that, with suitable redefinition of the field and 
taking particle limit and decoupling gravity, the prepotential 
 reduces to $SU(2)$ Seiberg-Witten prepotential. Let us 
denote new coordinate of moduli as $S=t_2=W_2/W_0,\ T=t_1=W_1/W_0$ 
where $S$ is heterotic dilaton, and 
near $y=0,\ x=1$, $S$ and $T$ behave as   
\bea
y=e^{-S}+\cdots,\ \ T={i\over 2\sin \la\pi}+\cdots.
\ena
Around $SU(2)$ enhancement point $T= i/(2\sin \la\pi)$, 
 we introduce new coordinate $\ti T$ 
as
\bea
\ti T=i{T-{i\over 2\sin \la\pi}\over T+{i\over 2\sin \la\pi}},
\ena
where such transformation is performed explicitly 
in appendix C. This choice is just a generalization of the observation made 
in ref.\cite{KKLMV}. Using these coordinates, prepotential around the 
conifold point can be written as
\bea
{\cal F}={1\over 2}S\ti T^2+{\cal F}_{1-loop}+{\cal F}_{nonpert}.
\ena
Putting fields as
\bea
&\ti T&=\sqrt{\alpha'}a,\ \ 
e^{-S}=\alpha'^2\La^4\exp(-\hat S),\\
&x_1&={x^2y\over (1-x)^2}={1\over u^2},\ \ x_2=1-x=\alpha'u, 
\ena
and taking particle limit $\alpha'\rightarrow 0$, 
 we can fix the period vectors $(1,S,\ti T,2F-S\pa_SF-\ti T\pa_{\ti T}F,
\pa_SF,\pa_{\ti T}F)$ completely by 
asymptotic form in the leading order, 
which satisfy the 
above constraint; 
\bea
{\cal F}=\alpha'{\cal F}_{SW}+O(\alpha'^2),
\ena
where ${\cal F}_{SW}$ is a prepotential of Seiberg-Witten theory in the 
weak coupling region. 
The specification of dual pair of fields in these models has been given in ref.\cite{KKLMV}. What we are going to show here is that the choice is consistent of the integrability of prepotentials by using the symplectic form constructed in the previous section. In terms of $Y_i$, this choice of basis 
can be written in the lowest order as 
\bea
&\ti z^0&\sim 1,\ \ \ti z^1\sim \ti T\sim x_2^{1\over 2},\ \ 
\ti z^2\sim S\sim \log(x_1x_2^2),\\
&\ti {\cal F}_0&\sim x_2\log(x_1x_2^2),\ \ 
\ti {\cal F}_1\sim x_2^{1\over 2}\log x_1,\ \ 
\ti {\cal F}_2\sim x_2.
\ena

Next let us calculate higher order terms beyond the 
Seiberg-Witten prepotential. Going to the higher 
order, there are some ambiguity 
since this system include several kinds of integer power solutions; 
for example, we can add $Y_2$ to $Y_0$ without changing 
leading behavior in this limit. 
That is, the combination of these solutions can not determined 
 by asymptotic behavior itself around the 
conifold point. 
This ambiguity is usually killed by the integrability of 
the prepotential. Without carrying out
tedious long calculation of this type we instead use the intersection form, 
since it is insured to be integrable by using 
the intersection form. 

As was the case with 
$1-$moduli, it is enough to consider 
solutions up to first few orders to calculate
 the intersection, because this is not changed 
if the order of calculation become higher. 
Using the intersection form (2.25) and 
changing the variable to $x_1,x_2$, 
we can give the intersection form which does not 
vanish as
\bea
C(Y_0,Y_5)={1\over 4},\ 
C(Y_1,Y_4)=-{1\over 8},\ C(Y_2,Y_3)={1\over 4}, 
\ena
even including higher order terms. From this 
form, we can deduce the combination of period vectors, 
however there are some ambiguity of linear combination of the solutions 
which does not change  value of intersection form. 
Now assuming proper asymptotic behaviors of $\alpha$ cycle
 $\ti z^0\sim 1,\ \ti z^1\sim 0,\ \ti z^2\sim \log y$, we fix 
this ambiguity in the basis.. There still remains 
an ambiguities, such as adding $Y_2$ to 
$Y_3$ as $Y_3'=Y_3+\beta Y_2$. The replacement is also consistent with the low energy behavior.  From the 
global consistency of moduli space as before, 
$\beta$ may be set to be zero. However we have not been able to fix the value
 at this moment. 

Thus we conclude the symplectic base around the conifold point 
to be 
\bea
&\ti z^0&\sim Y_0,\ \ti z^1\sim Y_1,\ \ti z^2\sim Y_3,\\
&\ti {\cal F}_0&\sim Y_5,\ \ti {\cal F}_1\sim Y_4,\ \ti {\cal F}_2\sim Y_2.
\ena
Using the definition 
of moduli as $\ti T\sim z^1/z^0,\ S\sim z^2/z^0$, 
and taking relative normalization of $Y_1$ and $Y_4$ as
$Y_4/Y_1=(1/\pi i)\log x_1+\cdots$ which obeys the physical 
constraint of realizing Seiberg-Witten theory\cite{SW} in the lowest order\cite{KKLMV}, 
 the prepotential of the theory is written by
\bea
{\cal F}&=&{1\over 2}{S\ti T^2}+
{1\over 2}\ti T^2\log \ti T^4+{1\over 2}\tau_0\ti T^2+
-{1\over 64}e^{-S}\ti T^{-4}\cdot \ti T^2
 -{5\over 32468}(e^{-S}\ti T^{-4})^2\cdot \ti T^2+\cdots \nonumber \\
& &+{-16+4\la-4\la^2\over 18}\ti T^4 -{
8-8\la+8\la^2\over 96}e^{-S}\ti T^{-4}\cdot \ti T^4
+\cdots  \nonumber\\
 & & +\cdots\\ 
&\rightarrow &\alpha'{\cal F}_{SW}+\alpha'^2\left\{
{-16+4\la-4\la^2\over 18} -{
8-8\la+8\la^2\over 96}{\La^4e^{-\hat S}\over a^4}\cdots \right\}
+\cdots  \nonumber
\ena
where higher order terms are listed in the appendix B. 

The degree of freedom of adding $\ti{\cal F}_2$ to $z^2$ 
is reflected to $\ti T^4$ term in the prepotential. 
On the contrary, the coefficient of the term $\ti T^{2n}\ (n\ge 3)$ does not 
contain $\beta$. This imply that 
 if $S$ independent terms of ${\cal F}$ excepting logarithmic term 
 is denoted as $h(T)$, which corresponds to the perturbative part,
 ${\pa_T}^5h(T)$ becomes true modular form. This is 
consistent with the result of the heterotic perturbation theory. 

\sect{Conclusion}
We have shown how to get the prepotential around the 
conifold point in type-II string theory compactified 
on various types of Calabi-Yau three fold up to 
$2-$moduli. We have introduced the intersection form which 
determine the symplectic form of period vectors of the 
solution of Picard-Fuchs equation. 
In $2-$moduli model, it is easy to give the the exact 
prepotential around the conifold point by 
using fixed periods, which is 
represented as the correction of 
the stringy effect from the Seiberg-Witten 
theory. 
Unfortunately we could not fix a parameter 
in prepotential around conifold locus even by 
using both duality conjecture and bi-linear 
forms. We may need full knowledge of analytic 
continuation, or direct instanton calculation on the heterotic side. 

Apart from the analysis of Calabi-Yau threefolds, we can obtain 
the bi-linear forms on Calabi-Yau d-folds\cite{GMP}. As a quick analysis 
of one moduli models\cite{NS,Sugiyama}, we find that even dimensional 
Calabi-Yau has symmetric forms contrary to the symplectic form in odd 
dimensions. These are consistent from the fact that intersection form 
of d-dimensional hypersurfaces are symmetric in even dimensions.

The treatment discussed in this article can be extended 
similarly to the other string theories, 
 whose gauge symmetry enhanced at conifold locus are known, such 
as 3-moduli models. Anyway the non-perturbative prepotential which 
is given by means of bi-linear forms around the conifold locus 
are not directly verified at the moment, 
since we do not know the formulation of direct calculation for 
the non-perturbative effect in the heterotic string theory. 
 We hope that some technique of the heterotic sting theory 
will be improved to estimate the justification 
of such macroscopic calculations some day. 
\appendix
\section{Normalization of periods in two-moduli models}
In large complex structure limit, 
the prepotential and Yukawa coupling of a series of models we considering 
 can be given by 
\bea
{\cal F}&=&-{1\over 3!}K_{ijk}t^it^jt^k+\cdots,\\
K_{111}&=&2K_{211}=2K_{121}=2K_{112}=8\sin^2 \la\pi.
\ena
Using Yukawa coupling and setting $\ti{D_{\rho_i}}={1\over 2\pi i}D_
{\rho_i}$, logarithmic  solutions 
and coordinates of moduli space $t_i$ can be written by \cite{HKTY1,HKTY2}
\bea
W_1&=&\ti D_{\rho_x}W_0,\ \ \ 
W_2=\ti D_{\rho_y}W_0,\\
W_3&=&-{1\over 2}K_{211}\ti D_{\rho_x}^2W_0,\ \ 
W_4=-{1\over 2}(2K_{121}\ti D_{\rho_x}\ti 
D_{\rho_y}+K_{111}\ti D_{\rho_x}^2) W_0,\\
W_5&=&{1\over 3!}(K_{111}\ti D_{\rho_x}^3+
3K_{112}\ti D_{\rho_x}^2\ti D_{\rho_y})W_0,\\
t_1&=&{W_1\over W_0},\ \ t_2={W_2\over W_0}.
\ena
With this normalization prescription, 
if we set the normalization of the 
intersection form as $\ti C=(K_{112}/(2\pi i)^3)C$, 
the intersection which does not vanish become 
$\ti C(W_0,W_5)=\ti C(W_1,W_4)=\ti C(W_2,W_3)=1$. 

If we use $\ti C$ around the conifold point to evaluate the intersection 
with taking the normalization of the solution as
\bea
Y_0&=&N_0(1+\cdots), \ Y_1=N_1(x_2^{1\over 2}+\cdots),\ 
Y_2=N_2(x_2+\cdots),\\
Y_3&=&N_3(\log(x_1x_2^2+\cdots),\ 
Y_4=N_4(x_2^{1\over 2}\log(x_1)+\cdots),\ 
Y_5=N_5(x_2\log(x_1x_2^2)+\cdots),
\ena
the condition $\ti C(Y_0,Y_5)=\ti C(Y_1,Y_4)=\ti C(Y_3,Y_2)=1$ can be
written as
\bea
N_0\,N_5={4K_{112}\over (2\pi i)^3},\ \ N_1\,N_4=
{-8K_{112}\over (2\pi i)^3},\ \ N_3\,N_2={-4K_{112}\over (2\pi i)^3}.
\ena
Notice that if we set heterotic dilaton as $S=t_2=(1/2\pi i)\log y+\cdots$, 
which is equivalent $N_3/N_0=1/(2\pi i)$, then $SU(2)$ Seiberg-Witten 
solution naturally arise because of the relation $N_4/N_1=1/(\pi i)$. 
\section{Prepotential around the point of enhanced gauge symmetries in two-moduli models}
The prepotential around $SU(2)$ enhancement point is given by 
\bea
{\cal F}={1\over 2}S\ti T^2+{\ti T^2\over 2}\log \ti T^4+
{1\over 2}\tau_0\ti T^2+h(\ti 
T)+{\cal F}_{n.p.},
\ena
where $\tau_0=\log 4-6$ is the bare coupling. 
Coefficients of T in $h(T)$ up to order $ T^{12}$ are given by

\begin{center}
\begin{tabular}{|c|l|}
\hline
$\ti T^4$&${1\over 18}(9b-4(4-\la+\la^2))$\\
\hline
$\ti T^6$&$-{2\over 675}(
1-14\la+48\la^2-68\la^3+34\la^4)$\\
\hline
$\ti T^8$&$
{2(1-2\la)^2\over 99225}(10-305\la+969\la^2-1328\la^3+664\la^4)$\\
\hline
$\ti T^{10}$&${-4\over 5\cdot 893025}(21-1638\la +17432\la^2-80480\la^3
+209596\la^4$\\
 & $-335180\la^5+331648\la^6-188504\la^7+
47126\la^8)$\\
\hline
$\ti T^{12}$&${2\over 3\cdot 25725625}(-883471770+22522670\la+
2343887\la^2-52343680\la^3+
20001415\la^4 $\\
 & $-631410\la^5+18364886\la^6-25083360\la^7+18173880\la^8-
7935360\la^9+1587072\la^{10})$\\
\hline
\end{tabular}
\end{center}

We give here non-perturbative part up to 5-th order from perturbative part. 
Non-perturbative part ${\cal F}_{n.p.}$ is decompose as
${\cal F}_{n.p.}=\sum_i{\cal F}_{n.p.}^i\ti T^{i+2}$.  
Setting 
$p=e^{-S}\ti T^{-4}$, each coefficient of $p$ in ${\cal 
F}_{n.p.}^{i}$ is given by

${\cal F}_{n.p.}^{0}$

\begin{tabular}{|c|l|c|l|}
\hline
$p$&$-{1\over 64}$&$p^4$&$-{1469\over 4\cdot 1073741824}$\\
\hline
$p^2$&$-{5\over 2\cdot 16384}$&$p^5$&$-{4471\over 5\cdot 34359738368}$\\
\hline
$p^3$&$-{3\over 54288}$ & & \\
\hline
\end{tabular}

\ 

${\cal F}_{n.p.}^{2}$

\begin{tabular}{|c|l|}
\hline
$p$&${-1\over 96}(-3\beta+8(1-\la+\la^2))$\\
\hline
$p^2$&${-1\over 2\cdot 4096}(-9\beta+4(4-\la+\la^2))$\\
\hline
$p^3$&${7\over 3\cdot 393216}(9\beta-4(4-\la+\la^2))$\\
\hline
$p^4$&$-{715\over 4\cdot 402953184}(-9\beta+4(4-\la+\la^2))$\\
\hline
\end{tabular}

\ 

${\cal F}_{n.p.}^{4}$

\begin{tabular}{|c|l|}
\hline
$p$&${1\over 1440}(-45\beta^2+240\beta(1-\la+\la^2)$\\
 & $-2(153-282\la
+404\la^2-244\la^3+122\la^4))$\\
\hline
$p^2$&${-1\over 2\cdot 18423}(117\beta^2-24\beta(20-11\la+\la^2)$\\
 & $+501-582\la+800\la^2-436\la^3+218\la^4)$\\
\hline
$p^3$&${-1\over 6\cdot 393216}(585\beta^2-720\beta(3-\la+\la^2)$\\
 & $+
2014-1436\la+1832\la^2-7-2\la^3-396\la^3+396\la^4)$\\
\hline
\end{tabular}

\ 

${\cal F}_{n.p.}^{6}$

\begin{tabular}{|c|l|}
\hline
$p$&${1\over 4\cdot 18900}(1575\beta^3-12600\beta^2(1-\la+\la^2)
+210\beta(153-282\la$\\
 &$+404\la^2-244\la^3+122\la^4)-4(
6590-16935\la+30789\la^2$\\
 & $-313274\la^3+24552\la^4-10698\la^5
+3566\la^6))$\\
\hline
$p^2$&${-1\over 967680}(-5355\beta^3+1260\beta^2(28-19\la+19\la^2)
-21\beta(3757$\\
 &$-5318\la+7536\la^2-4436\la^3+2218\la^4)+
60278-134739\la$\\
 &$+253751\la^2-278240\la^3+239660\la^4-12
0648\la^5+40216\la^6)$\\
\hline
\end{tabular}

\ 

${\cal F}_{n.p.}^{6}$

\begin{tabular}{|c|l|}
\hline
$p$&${-1\over 6\cdot 529200}(
33075\beta^4-352800\beta^3(1-\la+\la^2)+
8820\beta^2(153$\\
 & $-282\la+404\la^2-244\la^3+122\la^4)-
336\beta(6530-16935\la$\\
 &$+30789\la^2-31274\la^3+24552\la^4-10698\la^5
+3626\la^6)$\\
 & $+8(165627-532161\la+1158022\la^2-1568356\la^3$\\
 &$+1635352\la^4-
1188258\la^5+674168\la^6-238356\la^7+59586\la^8))$\\
\hline
\end{tabular}
\section{Analytic properties of periods in $K3$ and torus}
In this appendix, 
 let us discuss in detail about the analytic properties of solutions. A part of this appendix will overlap with some results given recently in 
ref.\cite{BDFPTPZ}.  As 
was discussed before, the difficulty of the analysis 
around the conifold point 
is the luck of the knowledge of the analytic continuation. 
However for the sake of $K3$ fibration, we may be able to 
continue three of six solutions which become power function
around the conifold point. To see this, let us see 
the solution with no logarithmic term of $y$ 
around large radius in the limit $y\rightarrow 0$. 
These functions are generalized hypergeometric function $_3F_2$: 
\bea
&W_0&={}_3F_2(\la,{1\over 2},1-\la;1,1,x)=\sum_{n}
{(\la)_n({1\over 2})_n(1-\la)_n\over n!n!n!}x^n,\\
&W_1&=D_{\rho_x}W_0,\ \ 
W_3={1\over 2}K_{211}(D_{\rho_x})^2W_0.
\ena
Some formula for the hypergeometric function 
make us possible to continue to the conifold point. 
First of all, we rewrite the solution as the product of two 
hypergeometric function\cite{HTF,LY,NS}
\bea
_3F_2(\la,{1\over 2},1-\la;1,1,x)=({}
_2F_1({\la\over 2},{1-\la\over 2};1;x))^2.
\ena
Next we are going to use the usual analytic continuation formula
. However at this stage, naive continuation gives wrong 
result because this process exceed the branch. To 
implement this, we use quadratic transformation to 
 rewrite the argument 
\bea
_2F_1({\la\over 2},{1-\la\over 2};1;x)={}_2F_1(\la,
1-\la;1;{1\over 2}-{1\over 2}(1-x)^{1\over 2}).
\ena
The formula of the analytic continuation 
of this type is well known. Therefore it is capable 
to continue $_3F_2$ function by using the representation of the 
product of two hypergeometric function
\bea
W_0=({}_2F_1(\la,
1-\la;1;{1\over 2}-{1\over 2}(1-x)^{1\over 2}))^2.
\ena
Similarly we can rewrite (D.18) as
\bea
W_1&=&{}_2F_1(\la,
1-\la;1;{1\over 2}-{1\over 2}(1-x)^{1\over 2})
\cdot D_{\rho_x}{}_2F_1(\la,
1-\la;1;{1\over 2}-{1\over 2}(1-x)^{1\over 2})\\
&=&{}_2F_1(\la,
1-\la;1;{1\over 2}-{1\over 2}(1-x)^{1\over 2})
\cdot {-1\over 2i\sin \la\pi}{}_2F_1(\la,
1-\la;1;{1\over 2}+{1\over 2}(1-x)^{1\over 2}),\nonumber \\
W_3&=&{K_{211}\over 2}\left\{
(D_{\rho_x}{}_2F_1(\la,
1-\la;1;{1\over 2}-{1\over 2}(1-x)^{1\over 2}))^2\right.\nonumber \\
& & \ \ \ \left.+{1\over 4\sin^2\la\pi}
({}_2F_1(\la,
1-\la;1;{1\over 2}-{1\over 2}(1-x)^{1\over 2}))^2\right\}\\ 
&=&({}_2F_1(\la,
1-\la;1;{1\over 2}+{1\over 2}(1-x)^{1\over 2}))^2+
({}_2F_1(\la,
1-\la;1;{1\over 2}-{1\over 2}(1-x)^{1\over 2}))^2.\nonumber 
\ena
Now we continue these solution to the conifold point. 
Define 
\bea
g_1(x)&=&{\Gamma({1\over 2})\over \Gamma({\la\over 2}+{1\over 2})\Gamma
(1-{\la\over 2})}{}_2F_1({\la\over 2},{1-\la\over 2};{1\over 2};
1-x),\\
g_2(x)&=&{\Gamma(-{1\over 2})\over \Gamma({\la\over 2})\Gamma
({1-\la\over 2})}(1-x)^{1\over 2}
{}_2F_1({\la+1\over 2},1-{\la\over 2};{1\over 2};
1-x),
\ena
and using the analytic continuation formula
\bea
{}_2F_1(\la,1-\la;1;{1\over 2}\pm{1\over 2}(1-x)^{1\over 2})
=g_1(x)\mp g_2(x),
\ena
results of the analytic continuation 
can be written 
in the form
\bea
W_0&=&(g_1(x)+g_2(x))^2,\\
W_1&=&{-1\over 2i\sin \la\pi}(g_1(x)-g_2(x))(g_1(x)+g_2(x)),\\
W_3&=&2(g_1(x)^2+g_2(x)^2).
\ena

The moduli parameter which is defined 
as the ratio of the two period is transformed 
 as follows,
\bea
T={W_1\over W_0}={i\over 2\sin \la \pi}{g_1(x)-g_2(x)\over g_1(x)+g_2(x)},
\ena
where near the conifold 
point, $g_1\sim 1$ and $g_2\sim 0$. 
This result is interpreted to the location of the gauge enhancement point 
in the moduli space
\bea
T={i\over 2\sin \la \pi}, 
\ena
which is the fixed point of the discrete subgroup 
of $SL(2,Z)$ of each model, especially if  $\la={1\over 6}$, this 
point is a fixed point of the transformation
\bea
T'=-{1\over T}.
\ena
Converting (3,30) to the expression for $g_1$ and $g_2$
\bea
{g_2(x)\over g_1(x)}=\,{T -{i\over 2\sin \la\pi}
\over T +{i\over 2\sin \la\pi}},
\ena
we see that the ratio $g_2/g_1=\ti T$ is to be taken as the 
moduli around the conifold point, which is 
just the redefinition of the field around the enhanced point of 
 gauge symmetry. 

Though it is difficult to 
handle the analytic continuation of the solution with $\log y$, 
without carrying out this, 
we are able to write down the symplectic transformation 
from the period in the large moduli limit to 
the one around the conifold point, by imposing this 
transformation must be symplectic. As is expected 
from the transformation law of $T$, this 
transformation does not reduce to 
$Sp(6;Z)$ from $Sp(6;C)$ in general. 
In the case with $\la=1/6$, 
if weak coupling behavior of heterotic dilaton 
is set to be $S=1/(4\pi i)\log y+\cdots$ rather than 
 the prescription in appendix A, and 
 moduli $T$ is set to absorb the factor $i$ as $T'=iT$, this 
transformation can reduce to $Sp(6;Z)$. 


As was pointed out in ref.\cite{KLM}, in the limit $y\rightarrow 0$ $K3$ moduli $\tau$ 
reduces to the moduli of a kind of torus, which is represented by 
\bea
\tau={i\over 2\sin \la\pi}{{}_2F_1(\la,
1-\la;1;1-z)
\over {}_2F_1(\la,
1-\la;1;z)},
\ena
where $z={1\over 2}-{1\over 2}(1-x)^{1\over 2}$, which we call 
$z_{1\over \la}$. This is a 
very similar form to usual moduli of the torus
\bea
\tau=i{{}_2F_1({1\over 2},{1\over 2};1;1-z_e)\over {}_2F_1(
{1\over 2},{1\over 2};1;z_e)},
\ena
however because of the factor $1/2\sin \la\pi$, the relation between 
$z_{1\over \la}$ and $z_e$ is complicated.  The relation to the absolute invariants are given in refs.\cite{KLRY,KLM,LY}. We will give the relation by using various transformation of hypergeometric functions given in 
refs.\cite{HTF,Goursat}. These identities has been used in the case of 
Seiberg-Witten theory\cite{MS}.

\begin{itemize}

\item $\la ={1\over 6}$;

On $\tau$ side, from quadratic transformation \cite{HTF}
\bea
{}_2F_1({1\over 6},{5\over 6};1;z_6)={}_2F_1({1\over 12},{5\over 12};1
;4z_6(1-z_6)).
\ena
On $\tau_0$ side, using quadratic and cubic transformation
\bea
{}_2F_1({1\over 2},{1\over 2};1,z_e)&=&
{}_2F_1({1\over 4},{1\over 4};1,4z_e(1-z_e))\nonumber \\
&=&
(1-z_e+z_e^2)^{-{1\over 4}}{}_2F_1(
{1\over 12},{5\over 12};1;{27z_e^2(1-z_e)^2\over 4(1-z_e+z_e^2)^3}
).
\ena
Thus the relation of $z_e$ and $z_6$ is \bea
4z_6(1-z_6)={27z_e^2(1-z_e)^2\over 4(1-z_e+z_e^2)^3} ={1\over J},
\ena
where $J$ is the absolute invariant\cite{HTF}. 

\item $\la={1\over 4}$;

On $\tau_o$ side, using quadratic transformation
\bea
{}_2F_1({1\over 2},{1\over 2};1;z_e)=(1-{1\over 2}z_e){}_2F_1
({1\over 4},{3\over 4};1;{z_e^2\over (2-z_e)^2}).
\ena
Thus $z_4$ ia relate to  $z_e$
\bea
z_{4}={z_e^2\over (2-z_e)^2}.
\ena

\item $\la={1\over 2}$; 

In this case, the kind of the function on both side are 
same initially. However this does not mean $z_e=z_2$ because 
of the factor $1/(2\sin \la \pi)$. In this case 
it is difficult to compare to each other, so,  
on  $\tau$ side, by 
using quadratic transformation 
\bea
{}_2F_1({1\over 2},{1\over 2};1;z_2)=
(1-z_2)(1+{4z_2\over (1-z_2)^2})^{-{1\over 2}}
{}_2F_1({1\over 4},{3\over 4};1;{4z_2\over (1+z_2)^2}),
\ena
we give the relation to $\la={1\over 4}$ case
\bea
z_4={4z_2\over (1+z_2)^2}.
\ena
Substituting this to (C.34) and solving for $z_2$ we read the relation 
to $z_e$ in $\la ={1\over 2}$ case as
\bea
z_2={z_e^2\over (1+(1-z_e)^{1\over 2})^4}.
\ena
\item $\la={1\over 3}$;
There seems to be any simple relations to other variables. However, we can find the relation to the absolute invariants by the quartic transformation\cite{Goursat}:
\bea
_2F_1({1\over 3},{2\over 3},1,z_3)=(1+8z_3)^{-{1\over 4}}\,_2F_1
({1\over 12},{5\over 12},1,{64z_3(1-z_3)^3\over (1+8z_3)^3}),
\ena
as
\bea
{1\over J} = {64z_3(1-z_3)^3\over (1+8z_3)^3},
\ena
as given in \cite{KLM,LY}
\end{itemize}

Unfortunately, we have not found the formula taking 
critical role in $\la={1\over 3}$ case.

\section{Intersection form in  K3 manifold}
In this appendix, we discuss briefly 
the possibility for applying 
our method by means of bi-linear form to  
different dimensional Calabi-Yau manifold. It is easy to 
expect that notion of symplectic intersection form 
directly acting on the solution space of the Picard-Fuchs equation, 
can be extended to arbitrary odd dimensional Calabi-Yau manifold. 
 Now we concentrate on the even dimensional case. In this case
 homology of the manifold consist of 
 $2n$-dimensional hypersurfaces, and its dual basis of 
cohomology are of $2n$-form. Therefore the interchange of 
these element appears to be symmetric. Taking 
into account of this situation, in spite 
of anti-symmetric operator (2.9), 
we way well construct symmetric bi-linear form by using 
anti-commuting operator
\bea
\{B,C\}(f_1,f_2)={1\over 2}(Bf_1\ Cf_2+Cf_1\ Bf_2). 
\ena

To be concrete, we are going to construct 
bi-linear form in the case of $K3$ manifold explicitly. 
Generally, the Picard-Fuchs operator of $K3$ manifold can be written by
\bea
D=\theta_x^3-x(\theta_x+\la)(\theta_x+{1\over 2})(\theta_x+1-\la).
\ena
Symmetric nature of cycles in this manifold is observed from 
the relation given by \cite{NS}
\bea
\int_{M}\alpha\wedge \beta=\int_{M}\beta\wedge \alpha=1,\ \ 
\int_{M}\gamma\wedge \gamma=2,
\ena
where $\alpha,\beta,\gamma$ are 2-forms corresponding 
to independent homology cycles. 
Now following the discussion of Section 2.1, 
we set bi-linear form $C$ close up to second order of $\theta$ by means of 
 operator (D.40) as
\bea
C=A_1\{1,\theta^2_x\}+A_2\{\theta_x,\theta_x\}+A_3\{1,\theta_x\}
+A_4\{1,1\}.
\ena
By requiring the condition that $C$ have to be constant 
\bea
\theta_x\ C=0,
\ena
we can find coefficients $A_i$ in following forms 
\bea
A_1=-2(1-x),\ A_2=1-x,\ A_3=x,\ A_4=x\la(1-\la). 
\ena
Immediately we can show that this bi-linear form 
can recover the previous result obtained in ref.\cite{NS} in the 
large radius limit. As a consequence, in the case of 
$K3$ space, method by means of bi-linear form 
can be used for specifying the period from the solution 
of Picard-Fuchs equation. 

Since it is easy to see that we can extend above analysis to 
the case of higher even dimension, we conclude that 
 the notion of bi-linear form can be 
employed in arbitrary even dimensional 
Calabi-Yau manifold, by introducing anti-commuting 
operator due to the symmetric nature of even 
dimensional homology cycle.


\newpage

\setlength{\baselineskip}{14pt}


\begin{thebibliography}{99}


\bibitem{Reid}
M. Reid, Math. Ann. {\bf 278},329 (1987).

\bibitem{CDLS}
P.Candelas, A.Dale, C.A.Lutken and R.Schimmrigk,
Nucl. Phys. {\bf B298}, 493(1988).

\bibitem{St1}
A. Strominger, 
Nucl. Phys. {\bf B451}, 96(1995).

\bibitem{GMS}
B.R. Greene, D.R. Morrison, and A.Strominger,
Nucl. Phys. {\bf B451}, 109 (1995).

\bibitem{CGGK}
T.M.Chiang, B.R.Greene, M.Gross and Y.Kanter, 
Nucl. Phys. {\bf B}, Proc. Suppl. {\bf 46}, 82(1996).

\bibitem{Greene}
B. R. Greene, ``{\it String Theory on Calabi-Yau Manifolds}'', hep-th/9702155.


\bibitem{KV}
S.Kachru and C.Vafa, 
Nucl. Phys. {\bf B450}, 69(1995).

\bibitem{KLM}
A.Klemm, W.Lerche and P.Mayr, Phys. Lett. {\bf B357}, 313(1995).

\bibitem{AL}
P.S.Aspinwall and J.Louis, 
Phys. Lett. {\bf B387}, 735(1996).

\bibitem{A}
P.S.Aspinwall, {\it K3 surfaces and string duality},hep-th/9611137.

\bibitem{CDFLL}
A.Ceresole, R. D'Auria, S.Ferrara, W.Lerche, and J.Louis,
Int. J. Mod. Phys. {\bf A8}, 79(1993).

\bibitem{KLT}
V.Kaplunovsky, J.Louis and S.Theisen, Phys. Lett. {\bf B357}, 71(1995).

\bibitem{HaM}
J.A.Harvey and G.Moore, 
Nucl. Phys. {\bf B463}(1996)315; Phys. Rev. {\bf D57}, 2329 (1998).

\bibitem{Koko}
C.Kokorelis, {\it The master equation for the 
prepotential}, hep-th/9802099.

\bibitem{CCLM}
G.L.Cardoso, G.Curio, D.Lust and T.Mohaupt,
Phys. Lett. {\bf B382}, 241(1996).

\bibitem{CCL}
G.L. Cardoso, G.Curio and D.Lust,
Nucl. Phys. {\bf B491}, 147 (1997).

\bibitem{HeM}
M.Henningson and G.Moore,
Nucl. Phys. {\bf B482}, 187(1996).

\bibitem{Kawai}
T. Kawai, Phys. Lett. {\bf B397}, 51(1997);
{\it K3 surfaces, Igusa cusp form and string theory}, hep-th/9710016; 
{\it String duality and enumeration of curves by Jacobi forms} hep-th/9804014.

\bibitem{C}
G.L.Cardoso, 
Nucl. Phys. Proc. Suppl. {\bf 56B}, 94(1997).

\bibitem{M}
G.Moore,{\it 
String duality, automorphic forms, and generalized Kac-Moody
   algebras} hep-th/9710198.

\bibitem{KKLMV}
S.Kachru, A.Klemm, W.Lerche, P.Mayr and C.Vafa, 
Nucl. Phys. {\bf B459}, 537 (1996).

\bibitem{KLMVW}
A.Klemm, W.Lerche, P.Mayr, C.Vafa and N.Warner, 
Nucl. Phys. {\bf B477} 746 (1996).

\bibitem{Lerche}
W.Lerche, {\it Introduction to Seiberg-Witten Theory and its 
Stringy Origin},\\ hep-th/9611190.

\bibitem{Kle}
A.Klemm, {\it On the Geometry behind N=2 Supersymmetric Effective 
Actions in Four Dimensions}, hep-th/9705131.

\bibitem{CPR}
P. Candelas, E.Perevalov, G.Rajesh, Nucl. Phys. {\bf B507}, 445(1997).

\bibitem{CdGP}
P.Candelas, X.de la Ossa, P.Green and L.Parks,
Nucl. Phys. {\bf B358}, 21(1991).

\bibitem{CdFKM}
P.Candelas, X.de la Ossa, A.Font, S.Kats, D.R.Morrison,
Nucl. Phys. {\bf B416}, 482(1994).

\bibitem{CFKM}
P.Candelas, A.Font, S.Kats, D.R.Morrison,
Nuc. Phys. {\bf B429}, 626(1994).

\bibitem{HKTY1}
S.Hosono, A.Klemm, S.Theisen and 
S.T.Yau, Comm. Math. Phys. {\bf 167}, 301(1995).

\bibitem{HKTY2}
S.Hosono, A.Klemm, S.Theisen and 
S.T.Yau, Nucl. Phys. {\bf B433}, 501 (1995).

\bibitem{B}
V.V. Batyrev, {\it Dual Polyhedra and Mirror Symmetry for Calabi-Yau
   Hypersurfaces in Toric Varieties}, alg-geom/9310003. 

\bibitem{BS}
V.V.Batyrev and D. van Straten,
Commun. Math. Phys. {\bf 168}, 493 (1995).


\bibitem{Suzuki1}
H.Suzuki, Int.J.Mod.Phys. {\bf A12}, 5123(1997).

\bibitem{Suzuki2}
H.Suzuki, Mod.Phys.Lett. {\bf A12}, 2847(1997).

\bibitem{SW}
N.Seiberg and E.Witten, Nucl. Phys. {\bf B426}, 19 (1994); 
Nucl. Phys. {\bf B431}, 484 (1994).

\bibitem{Griffiths}
P.Griffiths, Ann. Math. {\bf 90}, 460 (1969).

\bibitem{KT}
A.Klemm and S.Theisen, Nucl.Phys. {\bf B389}, 153 (1993).

\bibitem{LT}
A.Libgover and J.Teitelbaun, Duke Math. J.{\bf 69}(1),29, 
{\it Int. Math. Res. Notes}(1993).


\bibitem{H}
S. Hosono, {\it GKZ Systems, Gr\"obner Fans and Moduli Spaces of Calabi-Yau
   Hypersurfaces}'', alg-geom/9707003.

\bibitem{BFKS}
V.V.Batyrev, I.C.Fontaine, B.Kim and D. van Straten,
{\it  Conifold Transitions and Mirror Symmetry for Calabi-Yau
   Complete Intersections in Grassmannians}, alg-geom/9710022; {\it Mirror Symmetry and Toric Degenerations of Partial Flag Manifolds}, math/9803108.

\bibitem{HL}
S.Hosono and B.H.Lian, {\it GKZ Hypergeometric Systems and Applications to Mirror Symmetry}, hep-th/9602147.

\bibitem{No}
M.Noguchi, Int.J.Mod.Phys. {\bf A12}, 4973 (1997).

\bibitem{St2}
A. Strominger, 
Commun. Math. Phys. {\bf 133}, 163(1990).


\bibitem{LS}
M. Lynker and R.Schimmrigk,
Nucl. Phys. {\bf B484}, 562 (1997).

\bibitem{GMP}
B. R. Greene, D.R. Morrison and M.R. Plesser,
Commun.Math.Phys. {\bf 173}, 559 (1995).

\bibitem{NS}
M. Nagura and K. Sugiyama, Int.J.Mod.Phys.{\bf A10}, 233 (1995).

\bibitem{Sugiyama}
K. Sugiyama, Int.J.Mod.Phys. {\bf A11}, 229 (1996).

\bibitem{BDFPTPZ}
M.Billo, F.Denef, P. Fre, I.Pesando, W.Troost, A.van Proeyen and D. Zanon,
{\it The rigid limit in special K{\"{a}}hler geometry.} hep-th/9803228.


\bibitem{HTF}
see for example, A.Erd\'{e}lyi \it et al., ``Higher
Transcendal Functions'',\rm (McGraw-Hill, New York)
 Vol.\bf 1. \rm

\bibitem{KLRY}
A. Klemm, B.H. Lian, S.S.Roan and S.-T. Yau,
{\it A Note on ODEs from Mirror Symmetry}, hep-th/9407192.

\bibitem{LY}
B.Lian and S.T.Yau, Commun.Math.Phys. {\bf 176}, 163 (1996);
{\it Mirror Maps, Modular Relations and Hypergeometric Series I}, hep-th/9507151, {\it Mirror Maps, Modular Relations and Hypergeometric Series II},  hep-th/9507153.

\bibitem{Goursat}
E. Goursat, Ann. Sci. Ecole Norm. Sup.(2) {\bf 10}, 3 (1881).

\bibitem{MS}
T.Masuda and H.Suzuki,
Int.J.Mod.Phys. {\bf A12}, 3413(1997).

\end{thebibliography}
\end{document}